# Understanding the principles of pattern formation driven by Notch signaling by integrating experiments and theoretical models

Federico Bocci[1], José Nelson Onuchic[1,*], Mohit Kumar Jolly[2,*]

[1]Center for Theoretical Biological Physics, TX, USA
[2]Centre for BioSystems Science and Engineering, Indian Institute of Science, Bangalore, India

*Authors for correspondence:
José Nelson Onuchic: jonuchic@rice.edu
Mohit Kumar Jolly: mkjolly@iisc.ac.in

## Abstract

Notch signaling is an evolutionary conserved cell-cell communication pathway. Besides regulating cell-fate decisions at an individual cell level, Notch signaling coordinates the emergent spatiotemporal patterning in a tissue through ligand-receptor interactions among transmembrane molecules of neighboring cells, as seen in embryonic development, angiogenesis, or wound healing. Due to its ubiquitous nature, Notch signaling is also implicated in several aspects of cancer progression, including tumor angiogenesis, stemness of cancer cells and cellular invasion. Here, we review experimental and computational models to help understand the operating principles of cell patterning driven by Notch signaling. First, we discuss the basic mechanisms of spatial patterning via canonical lateral inhibition and lateral induction mechanisms, including examples from angiogenesis, inner ear development and cancer metastasis. Next, we analyze additional layers of complexity in the Notch pathway such as the effects of varying cell sizes and shapes, ligand-receptor binding within the same cell, variable binding affinity of different ligand/receptor subtypes, and filopodia. Finally, we discuss some recent evidence of mechanosensitivity in the Notch pathway in driving collective epithelial cell migration and cardiovascular morphogenesis.

# Introduction

Notch signaling is one of the most well-conserved transduction pathways in metazoans. Activation of Notch signaling consists of a well-conserved set of steps including binding, cleavage and transport of the Notch Intracellular Domain (NICD) and leads to consequent transcription regulation of a multitude of biological processes including cell differentiation, proliferation and death [1].

The Notch signaling cascade is initiated by the binding of an extracellular ligand to the transmembrane Notch receptor (Figure 1). Typically, the ligand is a transmembrane protein at the surface of a neighboring cell, but it can occasionally be a soluble ligand in the extracellular microenvironment [2]. The ligand-receptor binding and forces originated by endocytosis induce a conformational change in the structure of the Notch receptor. This modification exposes a region of the receptor that undergoes sequential cleavage by the enzymes ADAM and $\gamma$-secretase, hence resulting in the release of the Notch intra-cellular domain (NICD). The NICD translocates to the cell nucleus, where it regulates several target genes together with several transcriptional cofactors such as CSL and Mastermind (Mam) [1]. More details on the molecular details of the signaling cascade can be found in several excellent reviews [1,3,4].

Each of the steps involved in this intracellular signaling cascade raises unanswered questions that would improve our understanding of several developmental processes and may also provide key insights to alleviate several pathological conditions, including cancer [1,3–6]. Here, we explicitly focus on the role of Notch in coordinating cell fate decisions and patterning at a multicellular level, and how various experimental and computational models can be integrated to elucidate the underlying dynamical principles of pattern formation. Due to its multi-cellular nature, Notch signaling offers an opportunity to understand how cell-fate decision in individual cells may be relayed to generate emergent multi-cellular dynamics. Despite the simplicity of this design, different Notch ligands can orchestrate different principles of multicellular spatial patterning via different positive and negative feedback regulation between NICD and its transcriptional targets [7]. For instance, Notch signaling can coordinate a divergent cell fate between two neighboring cells, a process known as lateral inhibition [1]. Moreover, the Notch pathway can modulate the opposite process, the lateral induction [8,9], by coordinating a similar cell state among neighbors.

In this review, we offer a bird's eye view on how to interpret cell and tissue dynamics in biological systems with simple concepts such as lateral inhibition and lateral induction, discuss the limitations of these models, and highlight a novel set of questions that require integrating experimental investigations with concepts from quantitative mechanistic modeling. In doing so, we bring together the analysis of several biological systems as well as theoretical modeling approaches that have so far helped highlighting the emergence of common themes in the operating principles of the Notch pathway. For the sake of simplicity, technical details of the underlying biology and mathematical models have been occasionally omitted, and relevant literature has been suggested. Furthermore, given the extensive set of topics covered in this review, we have focused on certain experimental and/or theoretical models that are representative of a particular system, and point the interested readers to relevant reviews for in-depth discussions of specific areas of research.

In the first section, we analyze how Notch signaling can give rise to divergent cell fate – lateral inhibition – and convergent cell fate – lateral induction – among two neighboring cells. Experimental evidence and theoretical modeling have contributed to understanding the competition and synergy between these patterning mechanisms in various physiological and pathological systems, including angiogenesis, inner ear development and cancer metastasis. Moreover, we review the oscillatory dynamics of Notch signaling that can arise due to coupling with other signaling pathways, for instance, during somitogenesis. Further, in the second section, we have examined the role of various molecular and morphological features that introduce additional layers of complexity to the canonical Notch signaling outcomes. The scenarios discussed here include the role of cell shape and packing geometry, cis-interactions between molecules within the same cell, mechanisms that alter the binding affinity between ligand and receptor paralogs, and long-distance (beyond-nearest neighbor) signaling through filopodia. In the final section, we review evidence pointing to a role for mechanosensitivity in assisting Notch-driven cell-fate decision. Relevant examples discussed here include collective epithelial cell migration and cardiovascular morphogenesis.

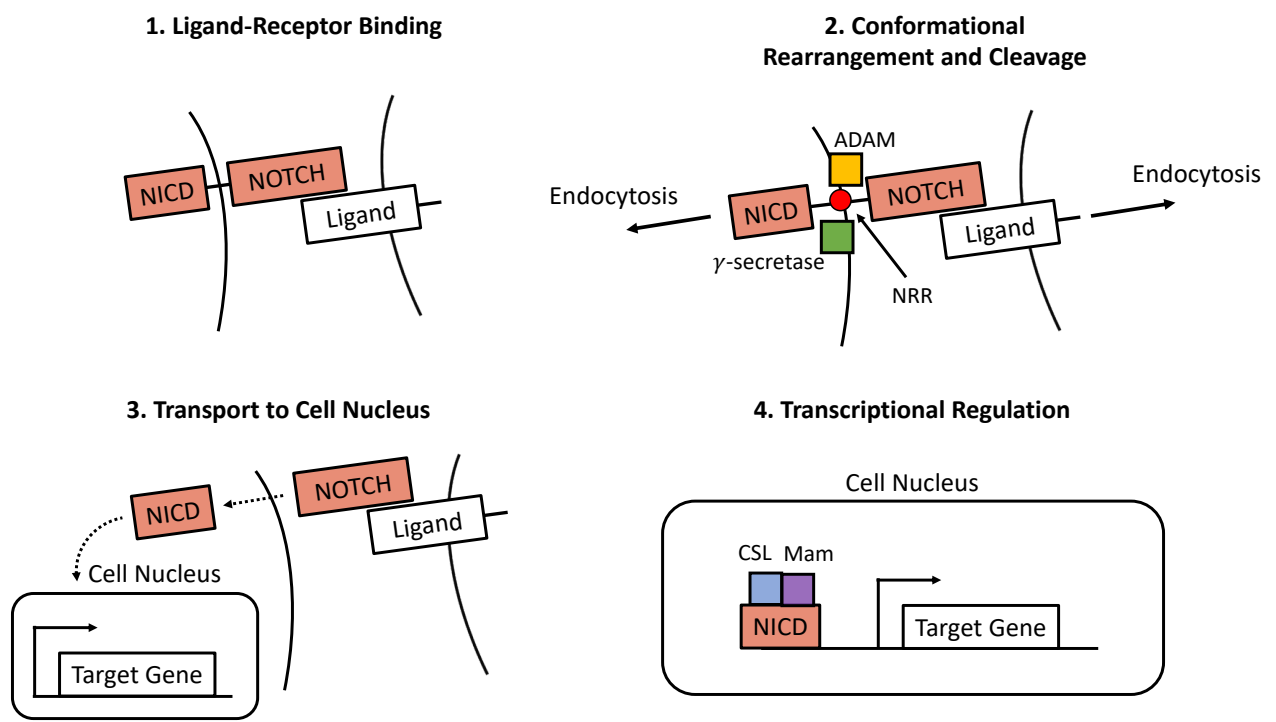

*Figure 1. **Intracellular and intercellular regulation through the Notch pathway.** The Notch transduction pathway. 1) Notch transmembrane receptor binds to a ligand at the surface of a neighbor cell. 2) Pulling forces originated in both cells expose the Negative Regulatory Region (NRR) of the receptor. Cleavages by ADAM and γ-secretase release the Notch Intracellular Domain (NICD) in the cytoplasm. 3) NICD is transported to the cell nucleus. 4) NICD transcriptionally regulates several target genes in cooperation with other co-activators such as CSL and Mastermind (Mam).*

# 1. Spatiotemporal patterning guided by Notch signaling

In this section, we review the experimental systems that exemplify two well-known patterning mechanisms enabled by Notch signaling: lateral inhibition and lateral induction. While lateral inhibition promotes opposite cell fates via biochemical negative feedbacks between the Notch receptor and ligands of the Delta family, lateral induction promotes similar cell fates by positive feedback between Notch and ligands of the Jagged family. Moreover, we review mathematical models that elucidate these patterning mechanisms on idealized, ordered lattices (such as the square and hexagonal lattices shown in Figure 3). Experiments and theoretical models help decoding the emergent outcomes of interactions between lateral inhibition and lateral induction mechanisms; specifically, we examine three biological processes that exhibit various degrees of patterning: angiogenesis, inner ear development and epithelial-mesenchymal transition in cancer metastasis. Lastly, we discuss temporal oscillations on Notch observed during somitogenesis as an example of spatiotemporal patterning.

**1.1 Biochemical mechanisms of lateral inhibition and lateral induction**

Historically, Notch signaling has been first characterized in *Drosophila melanogaster* as a mechanism that induces opposite cell fates among nearest neighbors [10–15]. The establishment of divergent phenotypes among two neighboring cells, or lateral inhibition, relies on binding of the Notch receptor to ligands of the Delta-like family (Delta in Drosophila; Dll1, Dll3 and Dll4 in mammals) presented at the cell surface of a neighboring cell [7,16]. Upon engaging of Delta with the transmembrane Notch receptor, the intra-cellular domain of Notch (NICD) is cleaved by enzymes and trans-locates to the cell nucleus. Here, NICD activates Hey/Hes1, which in turn inhibits Delta [1,4,17] (Figure 2A). This negative feedback amplifies small initial differences in ligand and receptor concentrations among nearly equivalent neighbors to establish opposite cell states. The cell with higher levels of Delta more effectively inhibits Delta in its neighbor, hence assuming a (low Notch, high Delta) phenotype, typically referred to as the Sender state, and forcing the neighbor to an opposite (high Notch, low Delta) phenotype, typically referred to as the Receiver state [18,19] (the green and orange cells in Figure 2A). This basic principle of differentiation regulates cell fate in several developmental and physiological processes. Interesting examples besides Drosophila's development include angiogenesis [20,21], spinal cord patterning in zebrafish [22–24], and development of neuroblast cells in early neurogenesis [25–27]. Thus, Notch-Delta system can be thought of as a two-cell 'toggle switch' [28] which can enable two opposite cell fates and possible switching among them under the influence of biological noise.

Despite being initially characterized as a driver of cell differentiation, Notch signaling can induce a convergent cell phenotype among neighbors through lateral induction [1,4]. A positive biochemical feedback between the Notch receptor and ligands of the Jagged/Serrate family establishes similar cell phenotypes that are spatially propagated to neighbors during the development of the inner ear [29–31] and vascular smooth muscle cell [32]. The Jagged family in mammals includes two paralogs (Jag1, Jag2), while Drosophila presents a single Serrate subtype [1,4]. Ligands of the Jagged/Serrate family are directly activated by NICD [32]. Therefore, Notch-Jagged signaling between neighbors can activate a positive biochemical feedback that establishes cell phenotypes with (high Notch, high Jagged) (the purple cells in Figure 2B), occasionally referred to as hybrid Sender/Receiver phenotypes to highlight that both cells send and receive signals [33]. Unless otherwise stated, green and orange colors denote high-Delta (Sender) and high-Notch (Receiver) phenotypes, respectively. Conversely, purple color indicate high-Jagged cells (hybrid Sender/Receiver) cells.

It is important to stress that positive and negative biochemical feedbacks that minimize or amplify initial differences are often assisted by a spatial and/or temporal regulation of Notch ligands and receptors (reviewed in more detail in [1]). For instance, in the development of the *D. melanogaster* wing imaginal disc, the ligand Serrate is expressed only by cells on the dorsal side due to spatial confinement of the upstream transcription factor Apterous [34]. This sharp boundary creates a stripe of Notch-active cells on the ventral side that leads to tissue growth thereafter [34].

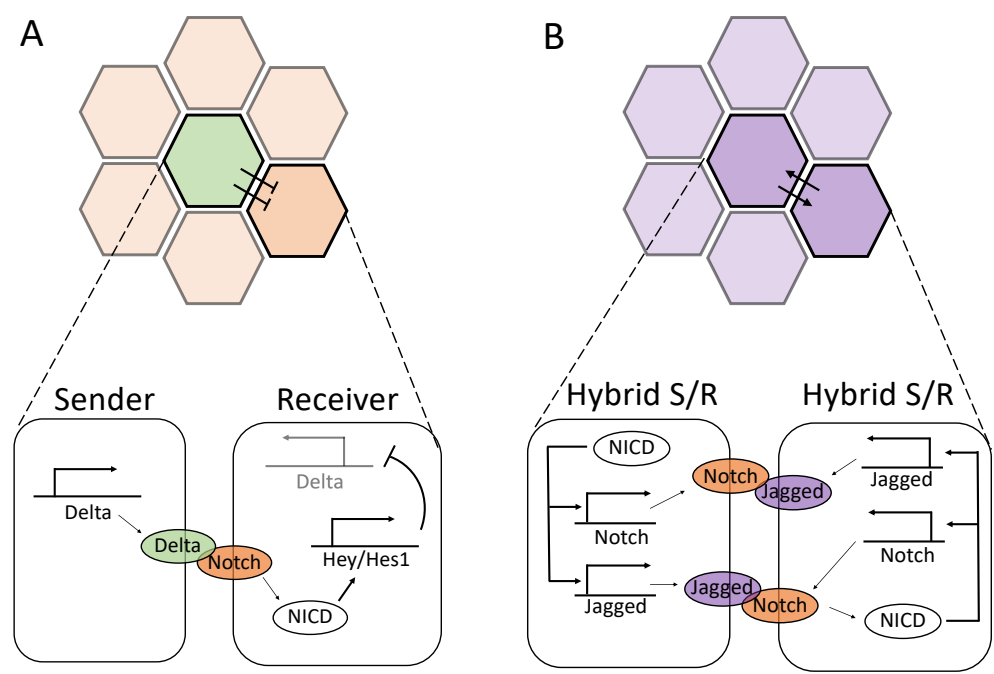

*Figure 2.* **Biochemical feedbacks giving rise to lateral inhibition and lateral induction in the Notch pathway.** *(A) In lateral inhibition, a high-expressing Delta Sender cell (green) suppresses the expression of Delta in its neighbors, hence enforcing a (low-Delta, high-Notch) or Receiver state. Detailed circuit: Delta ligands of the Sender cell activate Notch receptors in the Receiver. The released NICD activates Hey/Hes1, which in turn suppresses the production of Delta (pointed by the light shading of Delta promoter). Conversely, Notch receptors are not activated in the Sender cells; thus, Delta is freely expressed. (B) In lateral induction, neighboring cells mutually promote a similar hybrid Sender/Receiver state. Detailed circuit: upon activation of Notch receptors, NICD transcriptionally activates Notch and Jagged, hence establishing a high Notch, high Jagged hybrid Sender/Receiver state. In both panels, the color shading in the top highlights the two cells shown in the detailed circuit in the bottom.*

## 1.2 Theoretical exploration of the Notch-Delta-Jagged circuit

Over the last two decades, many theoretical models considerably helped understanding the biochemical feedback loops leading to lateral inhibition and lateral induction as well as the consequences of these signaling modes at the level of a cell population. In the first model of Notch-Delta lateral inhibition, Collier and colleagues [19] hypothesized that Delta's activation in a given cell stimulates Notch in neighboring cells, while Notch activation restricts Delta within the same cell (Figure 3A). In this model, the homogeneous state where neighbors express the same levels of Notch and Delta is found to be stable for weak biochemical feedback, while cells differentiate into a Sender and a Receiver for strong feedback [19]. When generalized to a spatial distribution of cells, cells tend to arrange in a 'salt-and-pepper' pattern where Senders are surrounded by Receivers and *vice versa* [19]. Therefore, cell patterning in the model depends on the geometric arrangement of cells. While Senders and Receivers can perfectly alternate on a square lattice, patterns on hexagonal lattices typically feature Senders surrounded by six Receivers, hence leading to a 3-to-1 Receiver/Sender ratio (Figure 3B). This type of patterning arises because a contact among Senders represent a more pronounced instability [35]. While a contact between Receiver cells results in the absence of signaling, two Sender cells in contact are likely to dynamically compete until one of them eventually become a Receiver [35]. This arrangement is well reflected, for example, in the avian inner ear, where hair cells are completely surrounded by supporting cells [36].

Further, some mathematical models have encapsulated the ability of Notch signaling to drive both divergent and convergent cell fates. A model developed by Boareto and colleagues considers the transcriptional activity of NICD that inhibits Delta and activates Jagged (Figure 3C). In this simplified representation, Delta and Jagged are generally representative of the two classes of ligand subtypes/families – one that is transcriptionally activated by NICD, the other one is repressed by NICD [33]. In this model, the positive feedback between Notch and Jagged can drive the cells away from lateral inhibition and promotes a convergent hybrid Sender/Receiver state. Therefore, if the relative contribution of Notch-Delta signaling is large as compared to that of Notch-Jagged, two neighboring cells fall into a divergent cell fate by lateral inhibition. If Notch-Jagged is dominant, however, the cells fall into a convergent 'hybrid Sender/Receiver' configuration with similarly high levels of Notch and Jagged [33]. Therefore, modulating the balance between Notch-Delta and Notch-Jagged signaling in the model leads to transition between salt-and-pepper patterns and homogeneous patterns (Figure 3D). This trend is reminiscent of dynamical behavior of an intracellular 'toggle switch' coupled with self-activation, where the relative strengths of mutual inhibition (similar to that seen for Notch-Delta) and self-activation (similar to the topology of Notch-Jagged) can drive different cell fates [37].

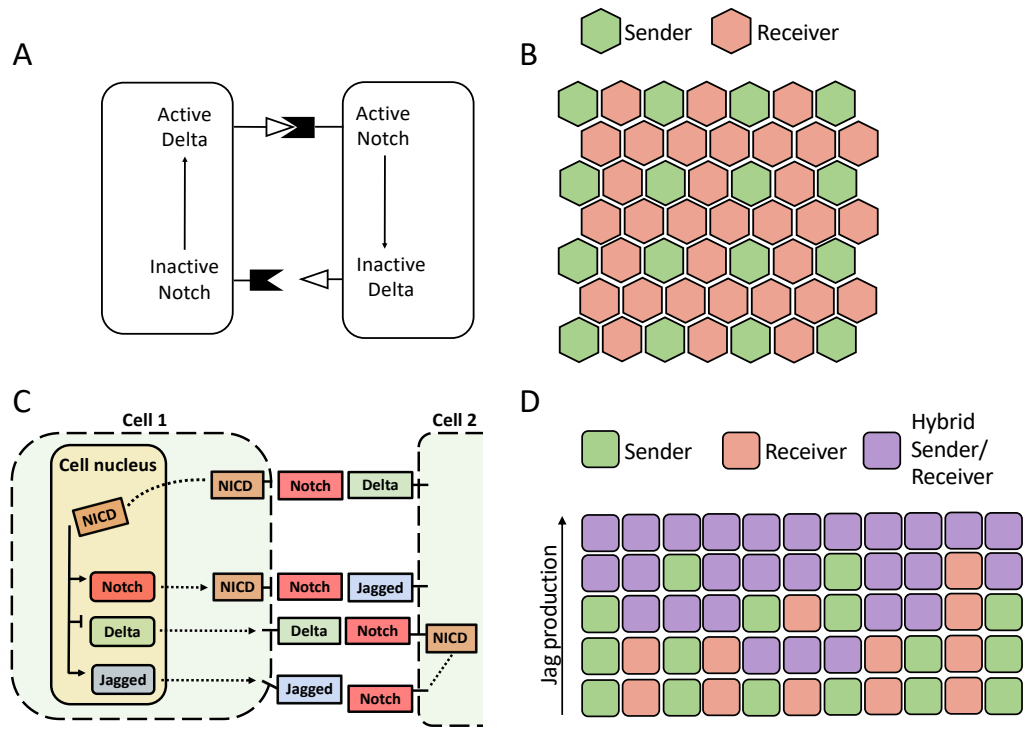

*Figure 3. Patterns predicted by models of Notch-Delta and Notch-Delta-Jagged signaling. (A) Schematic of the Notch-Delta cell-cell signaling model proposed by Collier and collaborators [19]. (B) A typical solution of the model of Collier and collaborators on a hexagonal lattice with Senders (green) surrounded by Receivers (red). (C) Model of the Notch-Delta-Jagged circuit proposed by Boareto and collaborators [33]. Solid black arrows in the cell nucleus indicate transricptional action of NICD. Dashed black lines indicate indicate transport of Notch, Delta and Jagged molecules to cell surface, where they can bind to ligands and receptors of a neighbor cell. (D) In the model of Notch-Delta-Jagged circuit, increasing the cellular production rate of Jagged destabilizes an alternate pattern of Senders and Receivers in favor of a homogeneous array of hybrid Sender/Receiver. Each row represents the pattern on a different one-dimensional chain of cells with increasing production rate of Jagged. Chains of cells with low production of Jagged show an alternation of Senders and Receivers, while chains with higher Jagged production rates show progressively more hybrid Sender/Receiver cells.*

## 1.3 Interplay of lateral inhibition and lateral induction described by experiments and mathematical models

Despite leading to opposite outcomes, lateral inhibition and lateral induction can take place at consecutive developmental steps, such as during inner ear development. Alternatively, they represent different outcomes that are selected based on signaling cues in the extracellular environment, such as during physiological or tumor angiogenesis. In this section, we review experiments and mathematical models that raise interesting questions about the interplay between lateral inhibition and lateral induction in three specific contexts: angiogenesis, inner ear development, and epithelial-mesenchymal transition during cancer metastasis.

*1.3.1 Angiogenesis*

Angiogenesis – the growth of new blood vessels from existing ones – is triggered by the hypoxia-induced signal VEGF (Vascular Endothelial Growth Factor). Secreted VEGF molecules bind to VEGF receptors (VEGFR) in the endothelial cells at the boundary of an existing blood vessel [21]. Activation of VEGFRs in turn leads to transcriptional activation of Dll4, hence inducing differentiation between a Tip cell with high Dll4, and a Stalk cells with low Dll4 by lateral inhibition [21,38]. Subsequently, tip cells develop filopodia and migrate toward the VEGF gradient, while stalk cells proliferate to support the formation of the new vessel (Figure 4A, top).

The model of Notch-Delta lateral inhibition in Tip-Stalk differentiation is supported by mathematical modeling of the VEGF-Notch-Delta signaling axis [39,40]. Moreover, computational models suggest that the Tip-Stalk selection process is highly kinetic, where the typical timescale to commit to a specific cell fate that vary considerably based on conditions in the extracellular environment as well as intracellular signaling dynamics [41].

A binary model of Tip-Stalk differentiation, however, cannot fully explain some experimental observations [21]. For instance, Dll4 can act as a brake on sprouting angiogenesis by inhibition endothelial tip formation [42]. Conversely, Jagged1 – which typically promotes lateral induction – promotes vessel development in mouse models where Notch-Dll4 signaling is antagonized by the glycosylation of Notch by Fringe [20]. In addition, lateral inhibition typically leads to patterns with alternate cell fates, while tip cells are typically separated by more than one stalk cell [20].

Various model shave been developed to explain deviations from classical Notch-Delta driven angiogenesis and the presence of partial Tip/Stalk states. Venkatraman and colleagues [41] showed that the regulators of Notch signaling such as lunatic fringe can significantly slow down the Tip-Stalk differentiation process, hence giving rise to metastable partial Tip/Stalk states [41]. To explain sparse patterns where Tips are separated by multiple Stalks, Koon and colleagues integrated a standard model of Notch-Delta lateral inhibition with intracellular heterogeneity of Notch concentration and tension-dependent binding rate of the Notch-Delta complex [43]. Interestingly, the addition of intracellular heterogeneity introduces states with intermediate levels of Notch and Delta, and gives rise to pattern with multiple cells in between consecutive Tips. Boareto and colleagues generalized their earlier computational model of the Notch-Delta-Jagged signaling circuit where external VEGF stimulus leads to bistability between the Tip phenotype (i.e. Sender) and the Stalk phenotype (i.e. Receiver) [44]. High expression of Jagged, however, stabilizes a homogeneous solution where cells assume a hybrid Tip/Stalk (i.e. hybrid Sender/Receiver) phenotype (Figure 4A, bottom). In this model's interpretation, lateral induction between hybrid Tip/Stalk cells can prevent a binary categorization of migrating and proliferating cells necessary for vessel development [44].

To elucidate the interplay between Dll4 and Jag1 during angiogenesis experimentally, Kang and colleagues exposed human endothelial cells to both VEGF signal and the pro-inflammatory cytokine Tumor Necrosis Factor (TNF) that activates Jag1 *in vitro* [45]. Strikingly, the combination of VEGF and low TNF dosage gives rise to longer vessels. At a critical threshold of TNF dosage, however, opposite outcomes (i.e. either robust vessel formation or no vessel formation) were observed in different experiments. Finally, TNF dosages above the critical dosage consistently prevented vessel formation [45]. Mathematical model focusing on the activation of Notch-Delta and Notch-Jagged signaling driven by VEGF and TNF, respectively, suggests a dose-dependent role for Jagged [45]. While high levels of Jagged can lead to hybrid Tip/Stalk cells, low Jagged levels acts synergistically with Dll1 to refine the alternate pattern of Sender and Receivers, hence contributing to more robust angiogenesis (Figure 4B). Therefore, increasing TNF dosage can lead to a switch in the role of Jagged from pro-angiogenesis to anti-angiogenesis [45].

In a pathological context, cancer cells can stimulate the sprouting of new blood vessels in the tumor microenvironment to supplement tumor growth [46,47]. Typically, tumors exhibit irregular vascular networks that prevent efficient drug delivery [48,49], and even facilitate passive metastasis by engulfing cancer cells [50,51]. The ability of cancer to induce vasculature makes tumor angiogenesis a potential therapeutic target to halt tumor progression. Strikingly, antitumor drugs that target Dll4, however, do not reduce tumor angiogenesis overall. Instead, anti-Dll4 drugs may result in a higher number of newly formed blood vessels with reduced functionality and chaotic architecture [47]. Lateral induction of the hybrid Tip/Stalk phenotype has been proposed as a potential explanation to this paradoxical finding. As anti-Dll4 drugs tilt the balance towards Notch-Jagged signaling, the lack of Tip-Stalk differentiation amplifying promiscuous cell differentiation and leaky angiogenesis [44].

As we gain a better understanding of the complex spatiotemporal dynamics of normal and tumor angiogenesis, the advantages and disadvantages of combining drugs targeting angiogenesis with other standard-of-care therapies demands further investigation. Limited exposure to vasculature can potentially protects the tumor from therapeutic agents that directly target cancer cells. Thus, perhaps counterintuitively, a transient renormalization of the tumor vasculature, timely synchronized with antitumor drugs, has been proposed as a potential strategy to alleviate tumor progression [52].

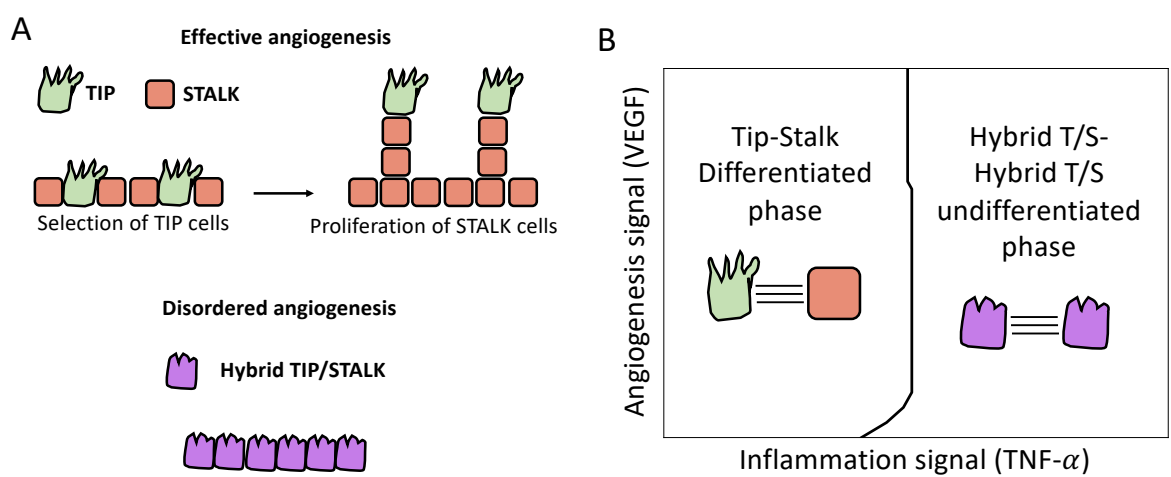

*Figure 4. **Physiological and pathological angiogenesis.** (A) Top: physiological angiogenesis is driven by cell differentiation between Tip (i.e. Sender, green) cells and Stalk (i.e. Receiver, orange) cells by Notch-Dll4 signaling. Bottom: lack of differentiation can lead to hybrid*

*Tip/Stalk cells (purple) and disordered angiogenesis as seen during tumor development.* ***(B)*** *In a model of two cells communicating via Notch-Delta-Jagged signaling, Kang and colleagues [45] predicted a transition from Tip-Stalk differentiation to hybrid T/S-hybrid T/S "de-differentiation" triggered by a threshold dose of TNF-α signal that activates Jagged.*

### 1.3.2 Inner ear development

Lateral induction and lateral inhibition operate progressively at different stages of the inner ear development to turn an initially homogeneous population of nonsensory cells into a refined mosaic of cells with specific phenotypes. The inner ear is composed by hair cells, that function as mechano-receptors that convert external stimuli into electrical signals, and supporting cells that provide tissue scaffolding, maintain a stable electrochemical environment, and occasionally differentiate to replenish the hair cell population after an injury [53,54]. During the prosensory cell specification phase, Notch activates Jag1, which in turn is capable of sustaining Notch in prosensory cells via a positive feedback (lateral induction) [55–57]. Thus, the activation of Notch and Jag1 not only establishes the hair cell phenotype, but also propagates it through lateral induction up to several cell diameters [8]. Later, in the hair cell differentiation phase, Notch-Dll1 signaling establishes the final pattern where hair cells (i.e. the Senders) are surrounded by supporting cells (i.e. the Receivers) [55–57]. For further insights on the role of Notch signaling in inner ear development, a thorough review is offered by Neves and colleagues [53]. Interestingly, Petrovic and colleagues argued with experiments and mathematical modeling that Jag1 acts synergistically with Dll1 during the hair cell differentiation phase in enforcing a robust lateral inhibition by acting as a competitive inhibitor for Dll1 [9]. Similar to the model of Notch-driven angiogenesis proposed by Kang and colleagues [45], a dose-dependent role for Jagged is suggested in inner ear development. While high levels of Jagged lead to a homogeneous state where cells attain a hybrid Sender/Receiver fate, a weak expression of Jagged can act synergistically with Dll1 to refine the alternate pattern of Sender and Receivers. In the presence of a dominant Notch-Delta signaling, additional Jagged tends to compete with Delta over binding Notch receptors, resulting in a greater activation of NICD, and thus suppression of Delta, in Receiver cells [9]. In this case, the ability of Jag1 to establish a convergent cell fate is negligible as compared to the cell differentiation promoted by Delta. When the signaling through the Notch-Jagged 'branch' of the pathway becomes too strong, however, lateral induction dominates the patterning (Figure 5) Interestingly, the dose-dependent role of Jagged is only observed in mathematical models of extended two-dimensional lattices. For instance, Boareto and colleagues [33] showed that Notch-Delta signaling robustly give rise to salt-and-pepper patterns of Sender and Receivers on a one-dimensional chain (see Figure 3D again). In the two-dimensional lattice cells have a higher number of nearest neighbors – and thus potentially contradictory external inputs to process – hence increasing the probability of mistakes, or Sender-Sender contacts, in the pattern.

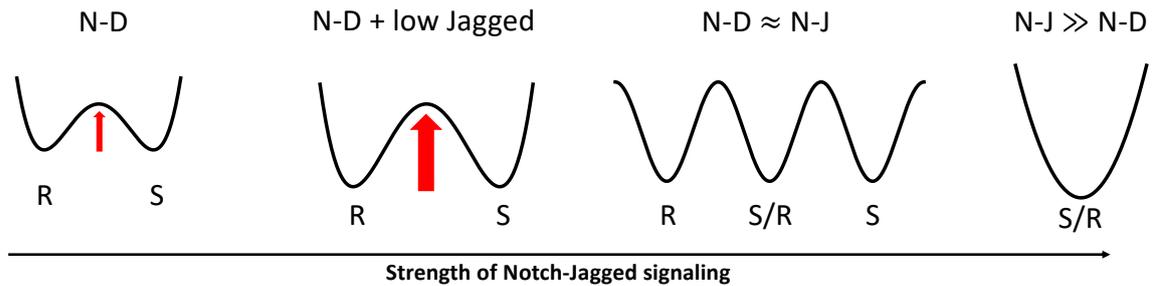

*Figure 5. **Proposed role of Jagged dosage in Notch-driven cell fate.** From left to right: in absence of Jagged (N-D), Sender and Receiver are the only accessible states in an abstract phenotypic landscape; a low Jagged dosage (N-D + low Jagged) increases the stability of Sender and Receiver states (indicated by the higher barrier in the landscape), as seen in inner ear development and angiogenesis; when both Notch-Delta and Notch-Jagged signaling are active (N-D≈N-J), a third hybrid Sender/Receiver state becomes accessible; an overwhelmingly strong Notch-Jagged signaling (N-J≫N-D) stabilizes the hybrid Sender/Receiver as the only accessible state.*

1.3.3 Epithelial-Mesenchymal Transition and cancer metastasis

Metastases represents the most critical step during tumor progression. Typically, cancer cells invade the circulatory system, reach anatomically distant sites and give rise to a secondary tumor [58]. These cells can migrate individually as well as collectively as multi-cellular clusters with varying size depending on cancer type, stage and patient individualities [59–61].

Generally, epithelial cancer cells partially or completely lose their cell-cell adhesion and acquire motility by undergoing the epithelial-mesenchymal transition (EMT) [62]. EMT can be activated by signaling cues in the tumor microenvironment in a cell autonomous manner as well as by Notch signaling. Activation of Notch signaling can be suppressed by EMT-inhibiting microRNAs such as miR-34 and miR-200 [63–66]. Notch signaling, however, can induce EMT by activating EMT-inducing transcription factors SNAI1/2 [67,68] (Figure 6A).

An effort to elucidate the coupled dynamics of Notch signaling with the EMT gene regulatory network [69] suggests that Delta-driven and Jagged-driven EMT can have different consequences at the level of multi-cellular patterning in a cancer tissue. While cells undergoing Notch-Delta-driven EMT are typically surrounded by epithelial cells, Notch-Jagged-driven EMT enables clustering among cells undergoing EMT (Figure 6B), hence potentially facilitating the formation of migrating multi-cellular cohorts in a tissue [69]. Besides, Jag1 can also stabilizing a hybrid epithelial/mesenchymal (E/M) cell phenotype [69]. Such hybrid E/M phenotype(s) can partially maintain cell-cell adhesion while gaining motility, and can invade as multi-cellular clusters that have elevated metastatic potential [59,70–72]. Experimental observations support this proposed role of Notch-Jagged signaling, although mostly through indirect evidence. First, CTC clusters from patients have a high expression of Jagged and co-express epithelial and mesenchymal markers, indicative of a hybrid epithelial/mesenchymal phenotype [60,73]. Conversely, single CTCs mostly lack

Jagged expression [60]. Second, Jag1 was identified as among top 5 differentially expressed genes in cells positive for K14, a marker for cluster-based migration [74]. Generalizations of this framework identified additional biochemical pathways that act as 'phenotypic stability factors' (PSFs) and stabilize hybrid E/M phenotype by coupling to the core Notch-EMT circuit. Examples include NUMB, NF-kB and IL-6 [75,76]. Consistently, overexpression of PFSs such as NUMB correlates with a worse patient survival in various cancer types [75–78].

To metastasize, migrating cancer cells need to proliferation potential and resistance to therapies typical of cancer stem cells (CSCs). Typically, cells undergoing a partial or complete EMT also show traits of CSCs [79–82]. Mathematical modeling of the gene regulatory networks underlying EMT, Notch and stemness suggests that Notch-Jagged signaling can promotes a 'window of opportunity' where cancer cells exist in a hybrid E/M, stem-like phenotype with aggravated metastatic potential [83,84]. Consistent with this prediction, CSCs display enhanced levels of Notch and Jagged across several cancer types including glioblastoma, pancreatic cancer, colon cancer and breast cancer [85–88]. Moreover, the glycosyltransferase Fringe which promotes Notch-Delta interactions over Notch-Jagged is reported as a tumor suppressor in multiple cancers [89–91]. Furthermore, it was recently shown *in vitro* that knockdown of Jag1 inhibits the formation of tumor emboli in hybrid E/M inflammatory breast cancer (IBC) - a rare but highly aggressive form of breast cancer that moves largely collectively through clusters [60] -  cells SUM149 [76].

Notch signaling can also regulate spatiotemporal pattern formation at the level of a tumor tissue. Analysis of breast cancer tissues highlighted subsets of mesenchymal CSCs at the tumor invasive edge, while subsets of hybrid E/M CSCs were largely localized in tumor interior [92]. A recent computational model developed by Bocci and colleagues suggests that Notch-Jagged signaling may contribute to generating this spatial heterogeneity. In the presence of a diffusive EMT-inducing signal such as TGF-β, Notch-Jagged signaling, but not Notch-Delta signaling, can give rise to large populations of CSCs. CSCs subsets at the tumor invasive edge are highly exposed to EMT-inducing signals and have a higher likelihood of undergoing EMT, whereas CSCs in the tumor interior are less exposed to EMT-inducing signals and hence retain a hybrid E/M phenotype [76]. Given the varying metabolic profiles of these CSC subsets [93], such patterning is reminiscent of spatial self-organization of metabolically diverse phenotypes in other contexts such as bacterial colonies [94,95].

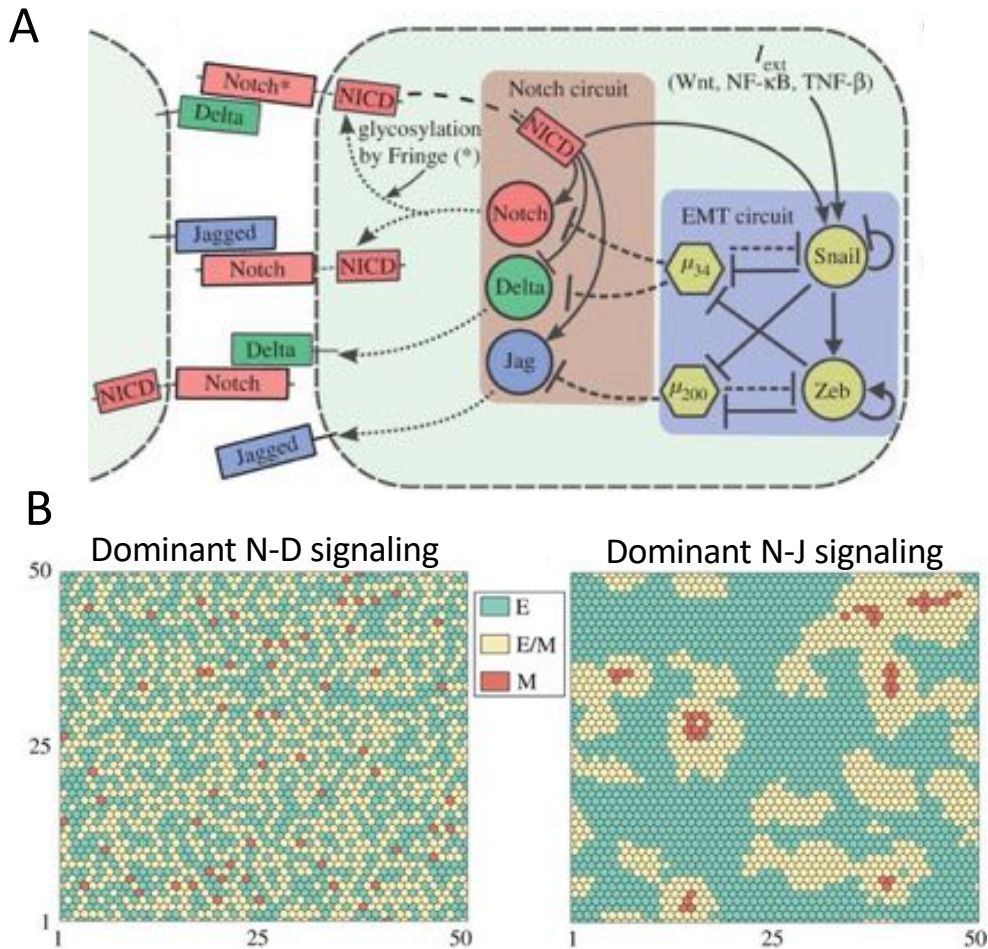

*Figure 6. **(A)** Proposed coupling between the Notch-Delta-Jagged circuit and the core EMT regulatory network proposed by Boareto and collaborators [69]. **(B)** Mathematical modeling of the Notch-EMT circuit predicts patterns where hybrid epithelial/mesenchymal and mesenchymal cells are mostly surrounded by epithelial cells in presence of dominant Notch-Delta signaling (left) and patterns with clusters of hybrid E/M cells in presence of dominant Notch-Jagged signaling. In this figure, green, yellow and red represent epithelial, hybrid epithelial/mesenchymal and mesenchymal cells, respectively. The figure is adapted from Boareto and collaborators [69].*

**1.4 Oscillations and synchronization as seen in the somite segmentation clock**

So far, we discussed mechanisms of spatial patterning. Due to its crosstalk with other signaling pathways, however, Notch can exhibit non-trivial temporal patterns. As an example, here we discuss somite segmentation, a well-known example of Notch oscillatory dynamics. During somite segmentation, the embryo's body axis is segmented into somites – blocks of epithelial cells that later give rise to vertebrae and tissues in the adult body [96]. Segmentation is organized by a precise spatiotemporal clock. Traveling waves of gene expression move along the body axis and stop at the location of a following segmentation event [96].

Oscillations in gene expression are generated in a cell autonomous manner via an autoregulatory negative feedback by Hes/Her proteins. Upon protein productions, Hes/Her molecules dimerize and suppress their own transcription [97,98]. The delay between transcription and protein synthesis gives rise to oscillations in Hes/Her gene expression (Figure 7) with a period of about 2-3 hours [99,100]. This model, however, is not sufficient to explain how oscillations maintain a precise cell to cell synchronization in time and space. Several experimental observations suggest a role for Notch-Delta signaling in synchronizing oscillations in neighboring cells, due to the biochemical coupling between the Notch and Hes/Her pathways. As previously discussed in section 3.1, NICD transcriptionally activates the family of Hes/Her molecules, which in turn, represses the expression of Delta [1,4,17]. Therefore, self-sustained oscillations of Hes/Her can potentially propagate to Notch (Figure 7). Zebrafish models indicate a periodic expression of Delta ligands during somite segmentation [101], while mouse models show oscillations of Notch, Delta and NICD [102–104].

Notch-Delta binding potentially provides information about the phase of the Hes/Her clock in neighbors. Mathematical modeling of the Notch-Hes/Her circuit developed by Lewis and colleagues [98,105] suggests that (1) oscillation can be self-sustained by the autoregulatory Hes/Her feedback loop, but (2) Notch-Delta progressively couples and eventually synchronizes the clocks of neighboring cells [98,105]. In other words, each cell can be viewed as an independent biochemical oscillator, and the exchange of ligands through the Notch receptor synchronizes the oscillations of the different cells [106] (Figure 7). This model is supported by observation in Zebrafish mutants that do not express Notch and Delta. In these mutants, segmentation is defective, and cells are arranged in heterogeneous patterns of high Hes/Her and low Hes/Her indicative of asynchrony in the cell population [107,108].

It remains unclear whether Notch's unique role is to ensure robust temporal correlation among neighbors. While it is generally accepted that Hes/Her self-inhibition is sufficient to generate temporal patterns, a number of studies in mouse models suggest that Notch might be required for oscillations. For further details, a comprehensive review on the role of Notch signaling in the somite segmentation clock is offered by Venzin and Oates [109].

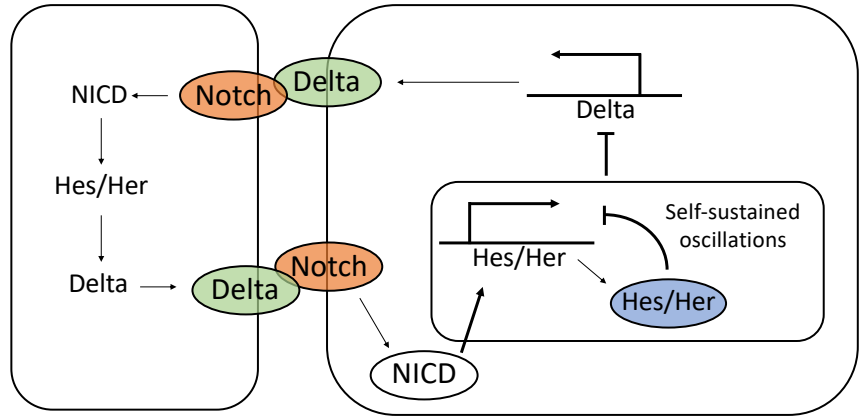

*Figure 7.* **The coupling between Notch-Delta and Hes/Her signaling synchronize temporal oscillations during somitogenesis.** Hes/Her can autonomously give rise to sustained oscillations by self-inhibition of Hes/Her protein. The coupling between Hes/Her and Notch-Delta signaling synchronizes oscillations between neighbors.

## 2. Non canonical modulation of Notch signaling

In the previous section, we discussed mechanisms of lateral inhibition and lateral induction guided by biochemical feedbacks between Notch and its ligands. In this section, we review mechanisms that modulate Notch signaling besides canonical positive and negative transcriptional feedbacks. These include dependence on cell-cell contact area and cell packing geometry, binding between receptors and ligands within the same cell, specificity in the affinity between receptor and ligand paralogs, and mechanisms enabling beyond nearest neighbor signaling. From a phenomenological standpoint, these mechanisms can be viewed as additional features beyond the simple nearest neighbor signaling introduced in the previous section.

### 2.1 Variability of cell packing and contact area

In the previous section, we developed a geometrical intuition on lateral inhibition that is based on alternate arrangement of Sender and Receiver cells. Mathematical modeling of Notch-Delta signaling helps understand these patterning dynamics on idealized ordered lattices. For instance, Notch-Delta signaling leads to a very specific pattern where Senders are surrounded by six Receivers on a perfect hexagonal lattice (see Figure 3B). Disordered lattices with variability in terms of cell size and number of nearest neighbors can lead to deviations from the standard 'salt-and-pepper' pattern.

The development of the basilar papilla, the avian equivalent of the mammals' organ of Corti, exemplifies how fluctuations in cell arrangement modulate lateral inhibition. The fully developed basilar papilla consists of a hexagonal mosaic where Sender cells (i.e. hair cells) are surrounded by six Receiver cells (i.e. supporting cells) [36]. Goodyear and Richardson found experimental evidence of dynamic cell rearrangement in the early development of the basilar papilla in a seminal study [36]. At earlier developmental stages (6-7 days), cell packing in the papilla is irregular. As a consequence of variable cell size and shape, the number of nearest neighbors fluctuate between 3 and 8 cells [36]. This underdeveloped mosaic allows occasional contacts between hair cells. Later on, cell packing relaxes toward a precise hexagonal mosaic and the "mistakes" in the patterning are corrected [36].

The size of shared contact area between neighbors is expected to fine-tune Notch signaling. Shaya and collaborators investigated the relation between cell size and cell fate by integrating experimental and computational methods [110]. By incorporating live-cell imaging reports to track the activity of Notch and Delta, they showed that signaling between pairs of nearest neighbors correlates with the cell-cell contact area. Smaller cells produced Delta at a higher rate and eventually became hair cells, while larger cells generally committed to a non-hair, supporting phenotype [110]. This result was reproduced by a mathematical model that generalized the seminal Notch-Delta lateral inhibition model [19] to a disordered lattice with variable cell sizes [110] (Figure 8). In the simplest model of lateral inhibition, Senders are selected from a homogeneous population by spontaneous breaking of symmetry and amplification of initial differences in protein levels [19]. Instead, this experiment shows that the fluctuations of cell size contribute to cell fate selection by introducing a weightage factor in the extent of Notch signaling between neighbors [110].

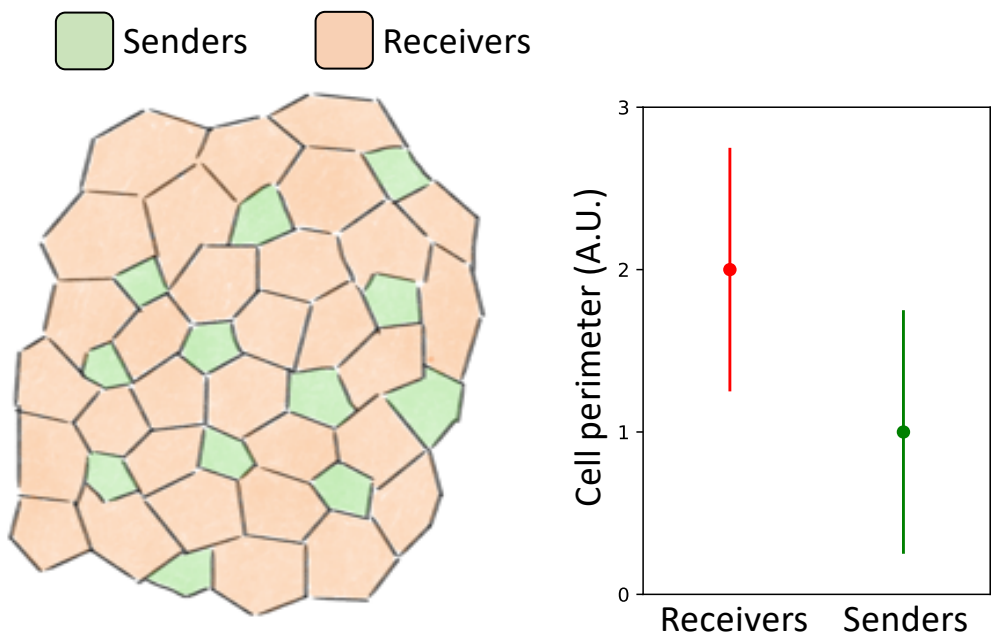

*Figure 8.* **Mathematical modeling predicts a correlation between cell size and fate.** *Mathematical model of Notch-Delta signaling on a disordered lattice developed by Shaya and colleagues [110] suggests that larger cells assume a Receiver phenotype and smaller cells assume a Sender phenotype. Left: a typical spatial patterning of Senders (green) and Receivers (red) predicted by mathematical modeling. Right: cells with large perimeter tend to become Receivers while cells with smaller perimeter tend to become Senders.*

## 2.2 Cis-interactions

Although Notch has evolved as a cell-cell signaling mechanism, receptors and ligands can bind within the same cell. Ligand-receptor binding within the same cell, or cis-interaction, does not lead to downstream signaling, but rather to ligand-receptor complex degradation, or cis-inhibition [14,111,112]. Despite not contributing to signaling, cis-inhibition can compete with the canonical Notch pathway by sequestering Notch receptors and ligands (Figure 9A).

Sprinzak and colleagues used time-lapse microscopy to evaluate Notch activation in response to external Delta ligands (standard trans-activation) and endogenous Delta (cis-interaction) [113]. While Notch receptors trans-activate gradually in response to external Delta, the response to indigenous, cis-Delta is sharp (Figure 9B). Therefore, Notch signaling is silenced when the level of intracellular Delta exceeds a threshold concentration [113]. This mechanism improves the robustness of lateral inhibition by further inactivating Notch in Sender cells. The authors further employed mathematical modeling to evaluate the behavior of an ensemble of kinetic models of Notch-Delta signaling with randomized parameters. Compared to a control model lacking cis-inhibition, the models that included cis-interactions yield lateral inhibition over a much broader parameter range by further refining defects in the patterning of Sender and Receiver cells [114]. The role of cis-inhibition, however, is not just restricted to proof-reading, but can rather be pivotal for cell-fate decision. For instance, loss of cis-inhibition compromises cell fate specification during the development of photoreceptors in Drosophila [115].

Although cis-interactions are mostly known to degrade Notch signaling without any contribution to signaling, experiments recently reported cell autonomous activation of Notch, such as in the cases of Drosophila bristle precursor cells and cell cycle regulation in T cells [116,117]. These experiments raise interesting questions about the competition between intracellular and intercellular signaling in modulating cell fate decisions. Nandagopal and colleagues engineered a synthetic system where cells constitutively express Notch while production of Delta is controlled experimentally [118]. Interestingly, extremal expression of Delta silenced Notch activity, whereas intermediate Delta expression maximized cis-activation [118]. To rationalize these observations, the authors developed various classes of mathematical models where cis-interactions can lead to either cis-activation or cis-inhibition with different rates. Interestingly, the non-monotonic response of Notch as a function of Delta concentration could only be reproduced by models with higher-order interactions and formation of clusters with multiple ligands and receptors [118]. Indeed, oligomerization of Notch receptors and ligands has been reported in the Notch pathway [119–121].

Given the role of cis-inhibition in enforcing robust lateral inhibition, it can be postulated that a switch from cis-inhibition to cis-activation would compromise precise cell patterns of Sender and Receiver cells. Formosa-Jordan and Ibanez [122] investigated the implication of Notch-Delta cis-activation in a disordered multicellular lattice model with variable cell size and shape. Compared to the mathematical model by Shaya and colleagues [110] discussed in the previous section, the authors did not focused explicitly on the correlation between cell size and cell fate, but rather on how cis-activation biases patterns of Senders and

Receivers. Their model confirms that cis-activation prevents robust lateral inhibition and introduces disordered patterns instead [122]. Specifically, for stronger cis-activation cell dynamics is predominantly cell-autonomous, rather than driven by nearest neighbors. Hence, cis-activation progressively increase the fraction of high-Delta Sender cells in the lattice model (Figure 9C). Indeed, cis-activation introduces a negative intracellular feedback where Delta ligands in the Sender cell promote their own inhibition by activating Notch receptors, hence driving the system away from the target Sender state with (low Notch, high Delta).

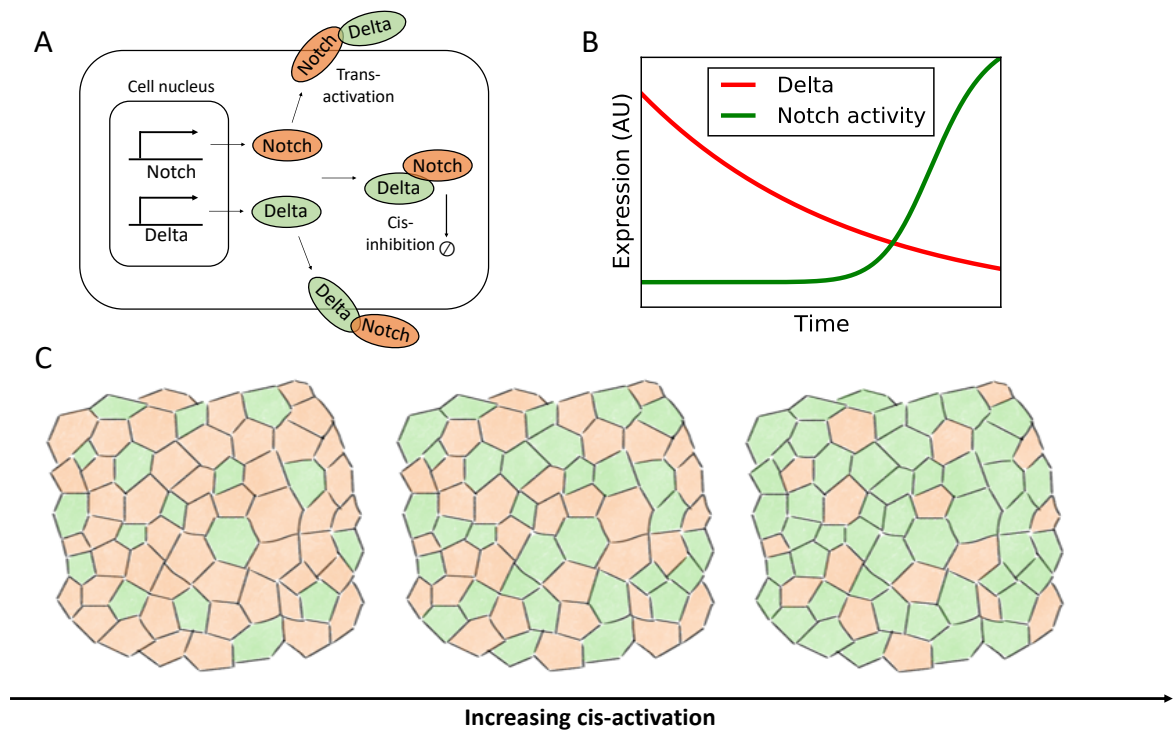

*Figure 9. **Cis-activation destabilizes the ordered lateral inhibition pattern.** (A) Binding of Notch and Delta molecules within the same cell leads to the degradation of the receptor-ligand complex without downstream signaling. (B) In a time-lapse microscopy experiment by Sprinzak and colleagues [113], the concentration of Delta (red) gradually decays exponentially due to dilution and cell division. Conversely, the activity of Notch (green) is turned on sharply when the concentration of Delta decreases below a threshold. This panel is adapted from Sprinzak and collaborators [113]. (C) In a model of Notch-Delta signaling on a disordered lattice developed by Formosa-Jordan and Ibanez [122], increasing the rate of cis-activation progressively disrupts lateral inhibition patterns. Left: in absence of cis-activation, Notch-Delta signaling gives rise to a pattern where Senders (green) are surrounded by Receivers (orange). For increasing levels of cis-activation, cell fate becomes cell autonomous and the fraction of Senders progressively increases (rightmost plots).*

## 2.3 Specificity in ligand-receptor binding affinity

The number of Notch receptor and ligand subtypes varies considerably in different species. Typically, mammals have four different paralogs of the Notch receptor (Notch1-4), three

Delta-like ligands (Dll1, Dll3, Dll4), and two Jagged ligands (Jag1, Jag2). Although the effect on the receiving cell is identical (i.e. NICD release), interactions through different ligand-receptor pairs can lead to differences in the downstream signaling cascade.

First, binding affinity varies based on the different molecular structure. For instance, Notch1 has a greater affinity to Dll4 than to Dll1 and Jag1 [123]. Moreover, different ligand-receptor pairings can lead to different dynamical responses in the receiving cell. For instance, Nandagopal and colleagues proposed that Notch1 can dynamically discriminate the ligands Dll1 and Dll4 in mouse and hamster cells [121]. Namely, while Dll4 activates Notch1 in a sustained manner, Dll1 gives rise to pulses of Notch1 activity [121]. Differences arise also in the ligand ability to cis-inhibit Notch receptors. For instance, Dll4 but not Dll1, can efficiently cis-inhibit Notch1 in mice cells [124], reminiscent of the greater Notch1-Dll4 affinity observed in trans-activation [123]. Moreover, the ligand Dll3 typically does not trans-activate any of the four Notch subtypes but only contributes to cis-inhibition [125,126].

Mechanisms that modify the binding affinity between the various subtypes of receptor and ligand can potentially result in a shift in cell fate by introducing an asymmetry between Delta and Jagged ligands. One such well-characterized mechanism is the glycosylation by Fringe proteins that results in a conformational change in the extra cellular domain of the Notch receptor [127,128]. Glycosylation typically decreases the binding of Notch with Jagged ligands both in trans- and cis-interactions [125,129–131]. Mathematical modeling of the Notch-Delta-Jagged signaling suggests that Fringe can stabilize the Sender and Receiver cell states by restricting the binding between Notch and Jagged, while loss of Fringe may tilt the balance towards lateral induction guided by Notch-Jagged signaling [37] (Figure 10A).

## 2.4 Interactions beyond nearest neighbor through filopodia

Although the Notch pathway is primarily designed as a pairwise interactions among nearest neighbors, beyond nearest neighbors' interactions are occasionally enabled by different mechanisms including filopodia and diffusible ligands.

Filopodia can extend up to several cell diameters and thus introduce contacts beyond nearest neighbor [132–134] (Figure 10B). For instance, in the bristle patterning of Drosophila, the sensory organ precursor cells (SOPs) with high Delta (i.e. the Sender cells) are separated by 4-5 receiver cells. This spacing, much larger than typically observed in lateral inhibition systems, is explained by dynamically rearranging filopodia that can give rise to transient contacts among non-neighbor cells [135]. This signaling between cells that are not adjacent to one another has been interpreted as a source of noise that refines the patterning [135].

Filopodia-driven signaling raises questions on how Notch can be effective when cells communicate through a small contact area. Khait and colleagues reported that the diffusion coefficient of Dll1 can vary over an order of magnitude (0.003-0.03 $\mu m^2/s$) from cell to cell in hamster ovary cells [136]. Based on this experimental finding, they developed a kinetic theoretical model including ligand-receptor binding at cell surface and lateral diffusion of Notch and Delta molecules across the cell surface. This framework highlights opposite regimes of signaling. When the size of the shared contact area between cells (b) is larger than the typical diffusion length scale ($\lambda$), diffusion effects are negligible and the signaling depends on the contact area. In the opposite regime ($\lambda > b$), however, the signaling strongly depends on the influx of Delta ligands in the contact area but only weakly on the size of the contact area itself [136] (Figure 10C). Diffusion coefficients in filopodia are larger by up to a 10-fold than in bulk membrane, possibly explaining how thin filopodia can still play an important role in Notch signaling [136].

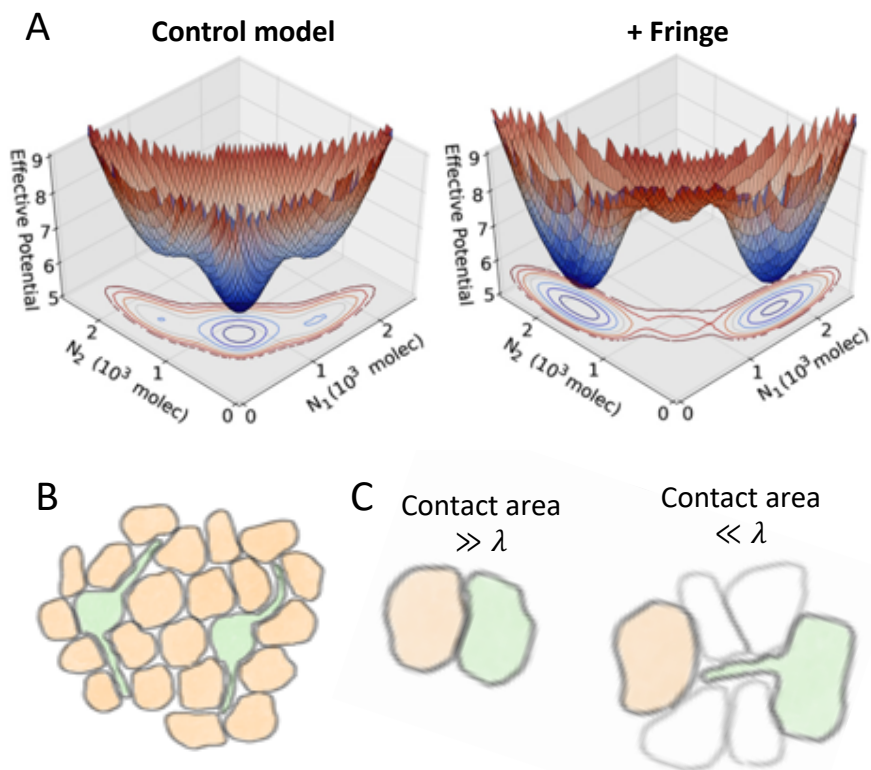

*Figure 10.* **Effect of Fringe glycosylation and filopodia on Notch signaling. (A)** *Mathematical model of Notch-Delta-Jagged signaling by Jolly and collaborators [37] predicts a switch in cell fate due to Fringe glycosylation. Effective potential of a two-cell model depicting the probability for the two cells to assume specific levels of Notch ($N_1$ and $N_2$, respectively). A control model without the effect of Fringe glycosylation (left) exhibit a single dominant minimum where both cells are hybrid Sender/Receiver with same Notch levels. Conversely, a model with Fringe modifies the landscape and introduces two separate states corresponding to Receiver-Sender ($N_1 \ll N_2$) and Sender-Receiver ($N_1 \gg N_2$). This panel is adapted from Jolly and collaborators [37].* **(B)** *Through filopodia, Sender cells (green) can potentially inhibit the Sender state in cells beyond nearest neighbors.* **(C)** *Schematic representation of the regimes of Notch-Delta signaling predicted by mathematical modeling by Khait and collaborators [136]. Left: when cells share a large contact area, diffusion of Delta ligands is negligible. Right: when the contact area is small, such as in the case of contact through filopodia, the signaling depends crucially on the diffusion of Delta ligands.*

# 3. Indications of a role for mechanosensitivity in Notch signaling

The activation of the Notch signaling cascade requires mechanical pulling on the ligand-receptor complex leading to NICD cleavage, and it therefore operates optimally within a certain range of mechanical constraints [137–139]. In contexts such as collective epithelial migration and cardiovascular morphogenesis, cells continuously modify their shape, tensions and stresses. It can be speculated that these biophysical factors add a further layer of regulation on Notch-driven cell patterning. While the role of mechanosensitivity is more quantitatively understood at the molecular scale of ligand-receptor interaction, its consequences at the level of multicellular patterning are still largely unexplored. The following two sections offer recently collected initial evidence of a role for mechanosensitivity in two processes driven by Notch: leader-follower differentiation during collective epithelial cell migration, and cardiovascular morphogenesis.

## 3.1 Lateral inhibition and mechanics select leader and follower cells during collective epithelial cell migration

Collective cell migration is commonly observed in physiological and pathological processes, including morphogenesis, wound healing and cancer metastasis. Collectively migrating cells conserve their cell-cell adhesion through several mechanisms, such as by maintaining adherens junctions [140,141]. Typically, some cells at the front of the migrating cell layer assume a distinct morphology characterized by an enlarged size and ruffling lamellopodia, and are labelled as 'leaders' at the migrating edge [142]. In a typical scratch assay that mimics wound healing, the mechanical injury at the boundary can generate a gradient of activation of several signaling pathways, with the strongest response in cells adjacent to the boundary and gradually decreasing in the inner region [143] (Figure 12A). Reminiscent of branching angiogenesis, the differentiation between leader and follower cells is regulated by the Notch-Delta pathway. Specifically, approximately 25% of the cells at the leading edge are leaders with high expression of Dll1. Conversely, cells with low Dll1 and high Notch become followers [144]. Interestingly, approximately 10% of cells transiently increase Dll1 after wounding but ultimately become followers, showing that the leader-follower differentiation is regulated in a highly dynamical manner by the Notch1-Dll1 pathway [144], similar to the dynamical balance of tip-stalk decision-making in angiogenesis [145].

Notably, leader-follower selection depends on feedback loops among Notch signaling and mechanical stresses. Indeed, receptor-ligand binding and the conformational change in the Notch1 domain thereafter require maintaining the receptor-ligand bond for enough time, which might be jeopardized by forces applied to the receptor or ligand [123,146], as can happen in the presence of mechanical injury during wound healing. Mechanical stresses inhibit the expression of Dll1 and prevent the selection of leader cells. Comparing the spatial distribution of mechanical forces and Dll1 expression suggests that the reduction of cellular stress at the boundary allows an effective Notch1-Dll1 signaling and leader-follower selection via lateral inhibition and gives rise to the observed gradient of Notch activation [144]. In the classic lateral inhibition scenario, Senders and Receivers are selected by stochastic fluctuations from competing cells that are initially in a similar cell state. A recent experiment showed via monolayer stress microscopy that mechanical interactions among followers cells behind the leading edge determine the selection and emergence of the leader cells at the leading edge [147]. In other words, this finding suggests that follower cells decide the leader, not the other way around as has been a long-held belief. Another recent study shows that a leader cell maintains its foremost spatial position for only a finite period of time; later, some followers can replace the leader cells that have consumed most of their energy, indicating a dynamic turnover or relay mechanism (Figure 12B) [148]. Such metabolic regulation is likely to be connected to Notch signaling; future investigations addressing the coupling between signaling, energy consumption and mechanics will be crucial to elucidate the dynamical principles of collective cell migration.

**3.2 Mechanosensitivity of Notch signaling in cardiovascular morphogenesis**

Evidence of Notch mechanosensitivity in leader-follower cell specification has been observed in a mouse model of retinal angiogenesis, where the Notch1-Dll4 pathway regulates the density of tip cells that give rise to new capillaries from the existing vasculature [149]. Although lateral inhibition is known to regulate tip/stalk differentiation during branching morphogenesis, this study showed that the tip/stalk differentiation heavily relies on the intercellular tension between cells in the blood vessel [149]. Similarly to observations in collective epithelial cell migration, tension between cells restricts Notch1-Dll4 signaling and compromises tip cell selection [149]. Overall, the density of tip cells and new branches was found to negatively correlate with the degree of mechanical stress, suggesting the Notch signaling might be tuned optimally at an intermediate intracellular tension that guarantees a proper angiogenic response, but limits the number of new branches [149]. Interestingly, intercellular tension regulates the Notch-Delta and Notch-Jagged pathways differently in the context of human cardiovascular morphogenesis. Laminar shear stress decreases the expression of Dll4 in human umbilical vein endothelial cells (HUVEC) – as observed in mouse angiogenesis – but also increases the expression of Jagged1, and overall potentiates the signaling between endothelial cells [150].

In the context of cardiovascular morphogenesis, the expression of Notch3, Jagged2 and multiple Notch targets decrease when a higher strain is imposed to vascular smooth muscle cells (VSMCs). Incorporating the dependence of Notch expression on strain into a computational model shows that the mechanosensitivity of Notch signaling is key in regulating the thickness of the vascular wall. In fact, a switch in cell patterning was observed in a model with an increasing number of VSMCs corresponding to the wall thickness [151]. For a short chain of cells (i.e. thin wall), most cells assumed a Sender state with high Delta. Conversely, think walls exhibited a chain of cells in a Sender/Receiver state with high Notch and Jagged levels [151].

The coupling between Notch signaling and mechanical forces is not unidirectional: Notch signaling can, in turn, regulate the function of vascular barriers that separate blood from tissues. For instance, Notch drives the assembly of adherens junctions in a non-canonical mechanism (i.e. not via transcriptional regulation of E-Cadherin levels) [152]. Consistently, reduction of Notch1 due to shear stresses leads to destabilization of adhesion junctions and proliferation of endothelial cells [153]. Therefore, Notch1 can potentially act as a mechanosensor by regulating the response of endothelial cells based on intercellular stresses, mechanical injuries, and angiogenic signals [153]. Therefore, while intercellular stresses might fine-tune Notch-Delta/Jagged signaling leading to new vessels, Notch signaling can, in turn, influence the defects in the structure of the vascular barrier by coordinating cell-cell adhesion. Future investigations integrating this interplay between biochemical aspects of Notch signaling, biomechanical aspects of mechanosensitivity, and the role of cell packing geometry will be valuable in elucidating the emergent dynamics of tissue-level pattern formation in different biological contexts.

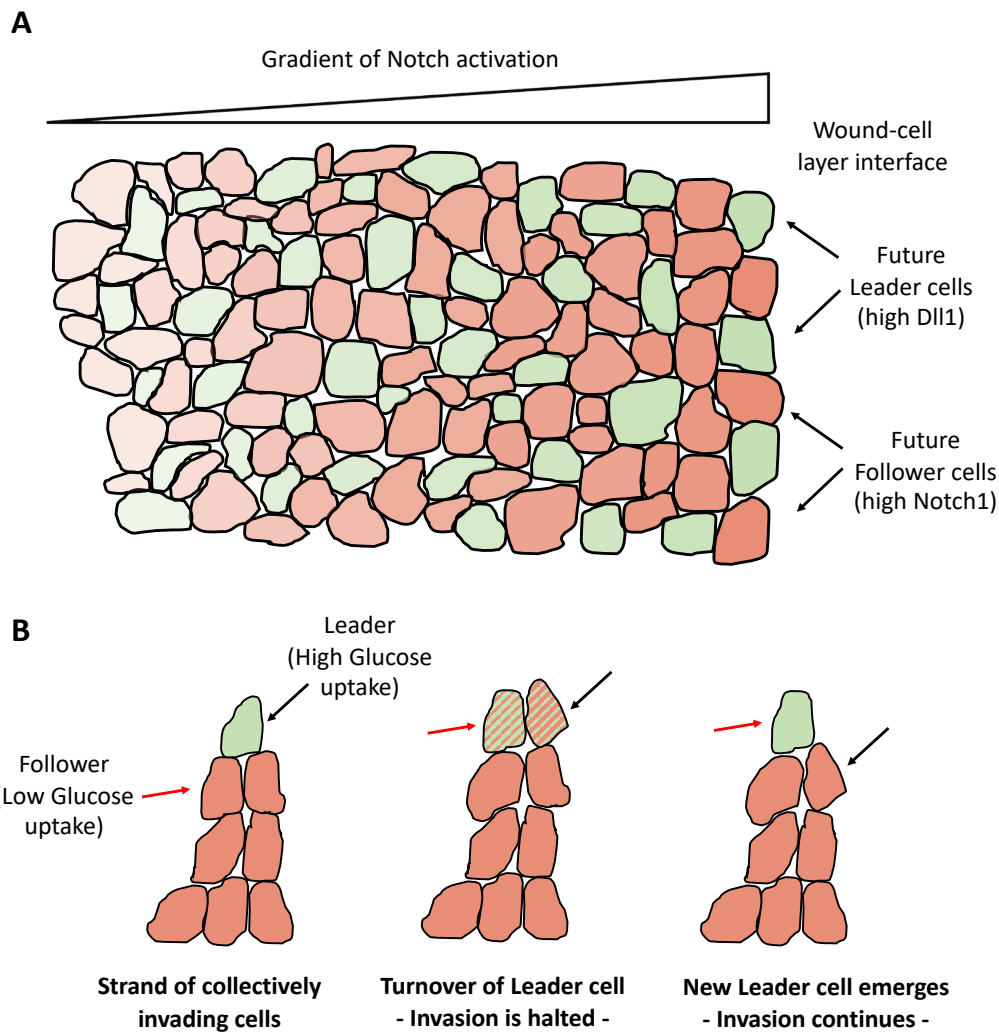

Figure 12. **Leader-Follower differentiation and turnover during collective epithelial migration. (A)** *Notch-Dll1 signaling differentiates cells that become leaders of the migration (green) and cells that become followers (orange). Notch-Dll1 signaling is more active toward the wound-layer interface (right end) and progressively inactivates far from the interface.* **(B)** *In strands of cells that migrate collectively, leaders have higher glucose uptake (left). Invasion halts in absence of a clear leader (center). The invasion continues after replacement of the leader cell (right). Red and black arrows highlight the two cells that exchange the leader position.*

## Open questions and future directions

Notch signaling is one of the most ubiquitous transduction pathways in vertebrates. Despite the variety of biological systems and processes, both physiological and pathological, that Notch signaling regulates, its structure and function are incredibly well-conserved.

Notch signaling has drawn incredible attention from the physics and mathematics community because, besides regulating cell-fate at a single cell level, it offers fertile ground to dissect the principles of spatiotemporal pattern formation in a tissue. To the eye of a physicist/ mathematician, Notch signaling gives rise to the modes of lateral inhibition and lateral induction similarly to a system of spins that align together or in opposition in a magnet. However, unlike magnetism, these different outcomes of cell states emerge from underlying molecular interactions that are often non-linear and can be separated in time-scale as well. The geometrical intuition about Notch patterning via lateral inhibition and lateral induction provides a key to interpret experimental observations in physiological processes such as embryonic development and angiogenesis [1,4,18]. For example, lateral inhibition correctly predicts alternate patterning where hair cells (i.e. Senders) are surrounded by supporting cells (i.e. Receivers) and make up about 25% of the total cell fraction, such as in the cases of inner ear development and collective cell migration [36,143]. Likewise, lateral induction describes well the propagation of similar cell fate observed, for instance, during inner ear development [9]. More investigations, however, will be needed to truly test how well these simple models of biochemical kinetics and feedback loops capture the signaling and patterning dynamics emerging from Notch at a quantitative level.

Moreover, most of the theoretical efforts toward understanding the operating principles of Notch have focused on deterministic models. Cell-to-cell variability, however, can arise due to both stochasticity in the intracellular biochemical signaling (intrinsic noise) and fluctuations of other cellular components and/or in the extracellular environment (extrinsic noise) [154]. Following a parallel between Notch and other patterning mechanisms driven by nearest neighbor signaling, such as the Ising model for a magnet, we speculate that stochastic fluctuation could play a relevant role in guiding, accelerating and/or disrupting ordered patterns [155].

Additional factors such as cell size and shape, affinity of ligand subtypes, molecular interactions within the same cell, and filopodia modulate the signaling. These mechanisms can be generally seen as details that add further complexity to the simple nearest neighbor's communication mechanism. For example, it is still not completely understood how trans- and cis-interactions integrate to establish cell fate. Cis-interactions between receptor and ligands of the same cell can typically lead to mutual degradation [14,111,112]. Recent evidence, however, suggests a role for cis-activation in the Notch pathway for multiple pairs of receptor and ligand subtypes [118]. Therefore, many context-specific signaling differences and their possible impact on spatiotemporal tissue dynamics deserve finer attention.

Moreover, early experimental findings suggest a role for mechanosensitivity in modulating Notch. The effects of extracellular forces on Notch activation are more quantified at the single molecule level [137–139]; it remains unclear, however, how these effects propagate at the level of multicellular patterning. On the experimental side, novel technologies that allow to probe the spatiotemporal Notch dynamics are starting to provide quantitative insights on the mechanochemical feedbacks between cell-cell signaling and cell mechanics [144,147]. On the other hand, integrating aspects of biochemical signaling, mechanical regulation and their interconnections is an important future challenge where theoretical and computational models can assist experimental design and *vice versa*.

Notch signaling has also received attention as a therapeutic target to curb cancer progression [5,6]. While theoretical modeling of signaling and regulatory dynamics typically adopts modular approaches that treat different signaling modules as independent blocks, Notch seems to be implicated in several hallmarks of cancer progression, including drug-resistance, leaky/chaotic angiogenesis and enhanced invasion and metastasis [5,6]. Jag1 is highly expressed in circulating tumor cell clusters with higher metastatic potential [60] and by cancer cells that resist to drugs [69,156] Generally speaking, cells that highly express Jagged seem to be associated with a more plastic and undifferentiated state such as hybrid epithelial/mesenchymal and/or a stem-like phenotype [76,85–88]. Therefore, quantifying the role of interconnections between Notch and other hallmarks of cancer invasion will be a crucial challenge at the crossing point between theoretical modeling, biology and data science.

Overall, insights from experimental and theoretical models continue to unravel the operating principles of Notch signaling, a master regulator of spatiotemporal cell patterning in development and tumor progression.

**Author contribution**
FB wrote the manuscript; JNO and MKJ guided the research and edited the manuscript.

**Conflict of interest**
The authors have no conflict of interest to disclose

**Acknowledgements**
The Onuchic Lab was supported by the National Science Foundation (NSF) grant for the Center for Theoretical Biological Physics NSF PHY-1427654, NSF grants PHY-1605817 and CHE-1614101. MKJ was supported by the Ramanujan Fellowship awarded by SERB, DST, Government of India (SB/S2/RJN-049/2018).